\documentclass[12pt]{article}
\usepackage{graphicx,amsmath}
\usepackage{color}

\newlength{\dinwidth}
\newlength{\dinmargin}
\begin{document}

\def\be{\begin{equation}}                                                    

\def\lapproxeq{\lower .7ex\hbox{$\;\stackrel{\textstyle                                                    
<}{\sim}\;$}}                                                    
\def\gapproxeq{\lower .7ex\hbox{$\;\stackrel{\textstyle                                                    
>}{\sim}\;$}}                                                    
\def\be{\begin{equation}}                                                    
\def\ee{\end{equation}}                                                    
\def\bea{\begin{eqnarray}}                                                    
\def\eea{\end{eqnarray}}

\begin{flushright}                                                    
IPPP/18/2\\
\today \\                                                    
\end{flushright} 

\vspace{1cm}

\begin{center}
{\Large\bf Black disk, maximal Odderon and unitarity}\\
\vspace{0.5cm}
 V.A. Khoze, A.D. Martin, M.G. Ryskin\\
 \vspace{1cm}

\begin{abstract} 

We argue that the so-called maximal Odderon contribution 
breaks the `black disk' behaviour of the asymptotic amplitude, since the cross section 
of the events with Large Rapidity Gaps grows faster than the total cross section. That is the `maximal Odderon' is not consistent with unitarity.\\
\end{abstract}
\end{center}
\vspace{0.5cm}
\section{Introduction}
 
 Recently the TOTEM collaboration at the LHC has published the results of the
first measurements at
 $\sqrt{s}$=13 TeV  of the  $pp$ total cross section $\sigma_{\rm tot}=110.6\pm 3.4$ mb 
\cite{Antchev:2017dia}
and of the ratio of the real-to-imaginary parts of the forward $pp$-amplitude, 
$\rho=$Re/Im$=0.10\pm 0.01$~\cite{TOT2}. 
Since the latter value appears to be  sufficiently smaller than that predicted by the
conventional COMPETE parametrization ($\rho=0.13 - 0.14$)~\cite{COMPET}, it may 
indicate either a slower increase of the total cross section at  higher energies or
a possible contribution of the odd-signature amplitude. (Note that within the COMPETE
parametrization the odd-signature term is described by secondary Reggeons and dies out
with energy.) Note that a $C$-odd amplitude, which arises from the so-called Odderon, and which depends weakly on energy,
is expected in perturbative  QCD~\footnote{QCD is the $SU(N=3)$ gauge theory which contains the spin=1 particle (gluon) and (for $N>2$) the symmetric colour tensor, $d^{abc}$. Due to these facts in perturbative QCD there {\em exists} a colourless $C$-odd $t$-channel state (formed from three gluons) with intercept, $\alpha_{\rm Odd}$, close to 1.}, 
see in particular \cite{Kwiecinski:1980wb,Bartels:1980pe,Bartels:1999yt}
and for reviews e.g. \cite{Braun:1998fs,Ewerz:2003xi}.
However the naive estimates show that its contribution is rather small;  
say, $\Delta\rho_{\rm Odd}\sim 1\mbox{mb}/\sigma_{\rm tot}\lapproxeq 0.01$~\cite{Rys} at the LHC
energies.
%is predicted by perturbative QCD (see for example the
%reviews~\cite{Braun:1998fs,Ewerz:2003xi}),

On the other hand, it is possible to introduce the Odderon phenomenologically 
%in the framework of asymptotoic theorems 
as an object which does not violate first principles and the axiomatic  theorems. 
In fact it was stated in \cite{Martynov:2017zjz} that the new TOTEM result is 
%  were advocated in \cite{Martynov:2017zjz} as
 a definitive confirmation of the experimental discovery
of the Odderon  in its maximal form.

Recall that the Odderon was first introduced in 1973~\cite{Lukaszuk:1973nt},
and since then it has been the subject of intensive theoretical
discussion, in particular within the context of QCD. Indeed, there have been several 
attempts to prove its existence experimentally
(see, for example,  
\cite{Braun:1998fs,Ewerz:2003xi,Block} for comprehensive reviews and references).
While the discovery of the long-awaited, but experimentally elusive, Odderon would be 
very welcome news for the theoretical community,
our aim here is to try to check whether the presence of the {\em maximal} Odderon,
with an amplitude $A^-\propto \ln^2 s$, which has a real part with high energy  
behaviour similar to that of the imaginary part of the even-signature amplitude
 $A^+$, does not violate unitarity at asymptotically large c.m.s. energy 
$\sqrt s \to\infty$.  Here we use the normalization $\mbox{Im}A=\sigma_{\rm tot}$.

\setcounter{figure}{1}
\section{Multi-Reggeon processes}
\vspace{-6.cm}
\begin{minipage}[c]{0.5\textwidth}
It was recognized already in the 1960s~\cite{VK-T,FK} that multi-Reggeon reactions, 
\be
pp\to p+X_1+X_2+...+X_n+p,
\ee
where small groups of particles ($X_i$), are separated 
from each other by Large Rapidity Gaps (LRG) (see Fig.1), may cause a problem 
with unitarity. Indeed, being summed over $n$ and integrated over the rapidities 
of each group, the cross section of such quasi-diffractive production increases 
faster than a power of $s$. This was  termed  in the literature
as the  Finkelstein-Kajantie  disease (FK),
see \cite{Abarbanel:1975me} for a review.
%This was called the Finkelstein-Kajantie (FK) problem.
\end{minipage}
\begin{minipage}{0.4\textwidth}
%Let us explain the situation using the simple example of Central Exclusive %Production of only one group/particle (Fig.2).....
%\end{minipage}
%\begin{figure}
\hspace{-.1cm}
\vspace{5.5cm}
\includegraphics[width=12cm]{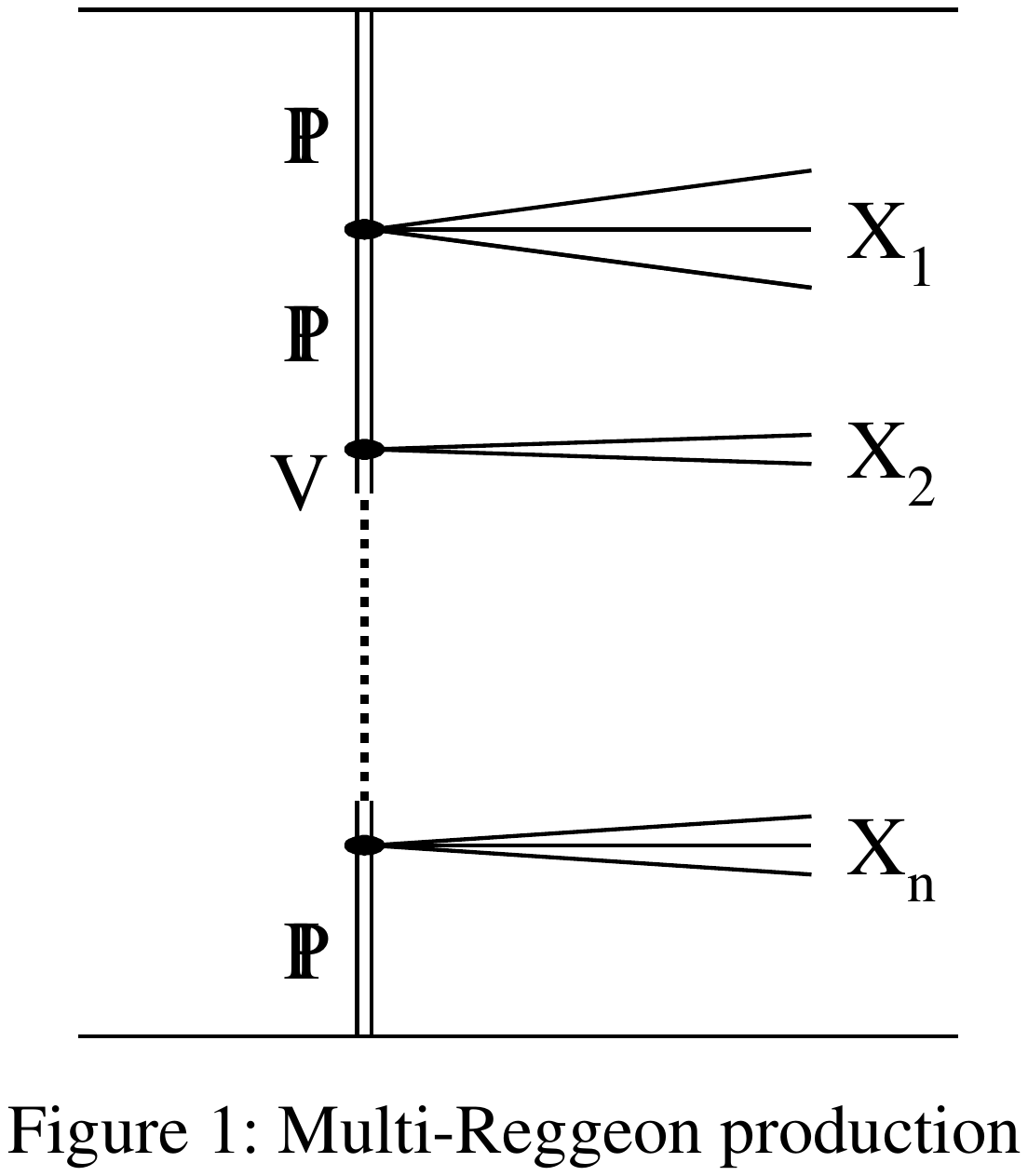}
%\caption{\sf Multi-Reggeon production}.
%\end{figure}
\end{minipage}

\vspace{-4.75cm}
Let us explain the situation using the simple example of Central Exclusive 
Production (CEP) of only one group of particles, as shown in Fig.2. Here the double line 
denotes the amplitude, $A$, which describes the interaction across the LRG 
(in particular, the proton-proton elastic amplitude).  Correspondingly, the CEP 
amplitude for Fig.2a reads
\begin{equation}
\label{a1}
A^{\rm CEP}(y_1,y_2,t_1,t_2)~=~A(y_1,t_1)\cdot V\cdot A(y_2-y_1,t_2)\ , 
\end{equation}
where $V$ is the vertex factor of central production and the $y_i$ are the values of the rapidity.

The full CEP cross section is given by the integral
\begin{equation}
\label{t-int}
\sigma^{\rm CEP}~=~N\int_0^Y dy_1\int dt_1 dt_2 ~
|A(y_1,t_1)\cdot V\cdot A(Y-y_1,t_2)|^2\ ,
\end{equation}
where $N$ is the normalization constant and where we put the upper rapidity $y_2=Y=\ln s$.
 \begin{figure}%[htb]
 \vspace{-5cm}
\includegraphics[scale=0.47]{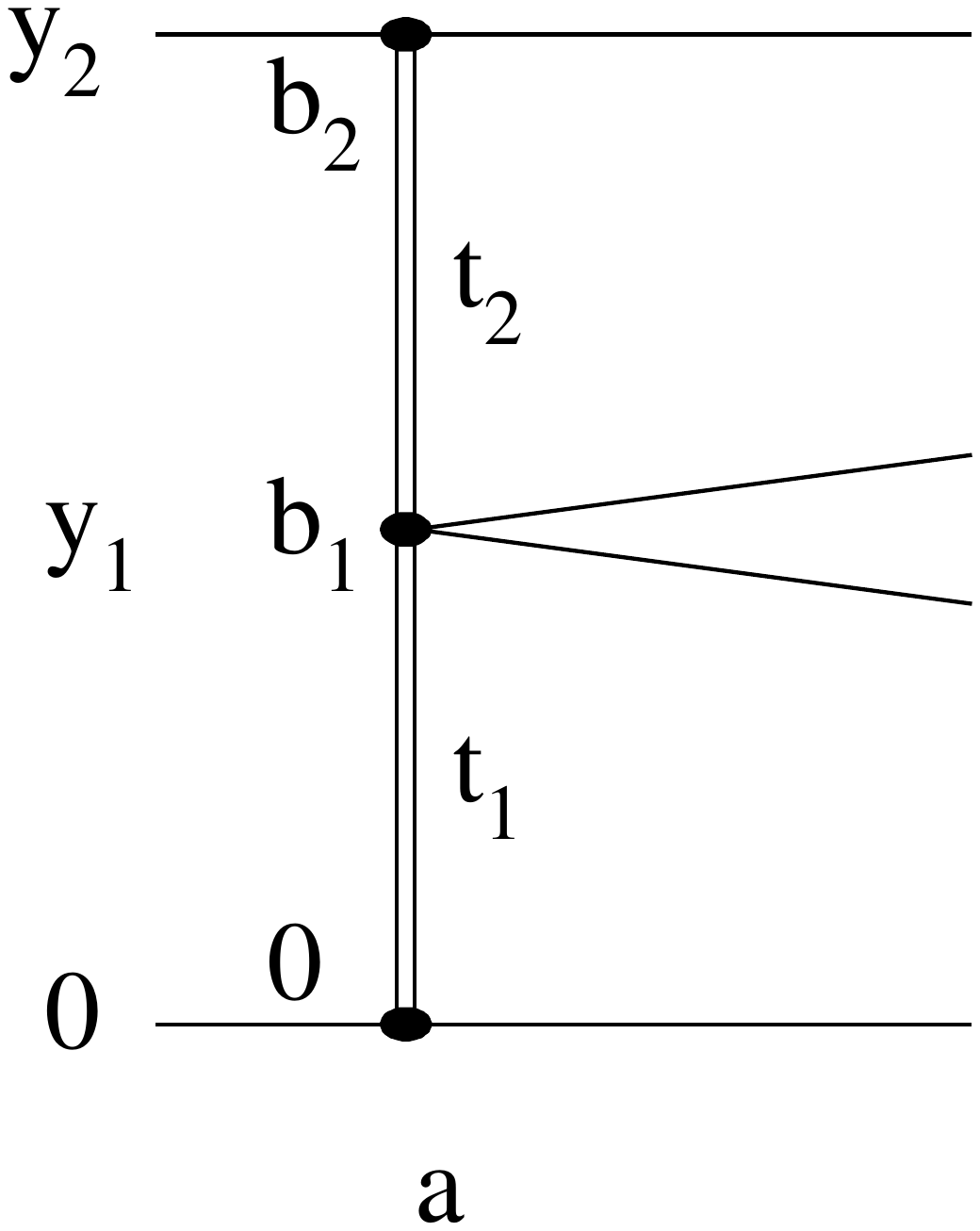}
\hspace{-3.3cm}
%\hspace{-4.3cm}
\includegraphics[scale=0.47]{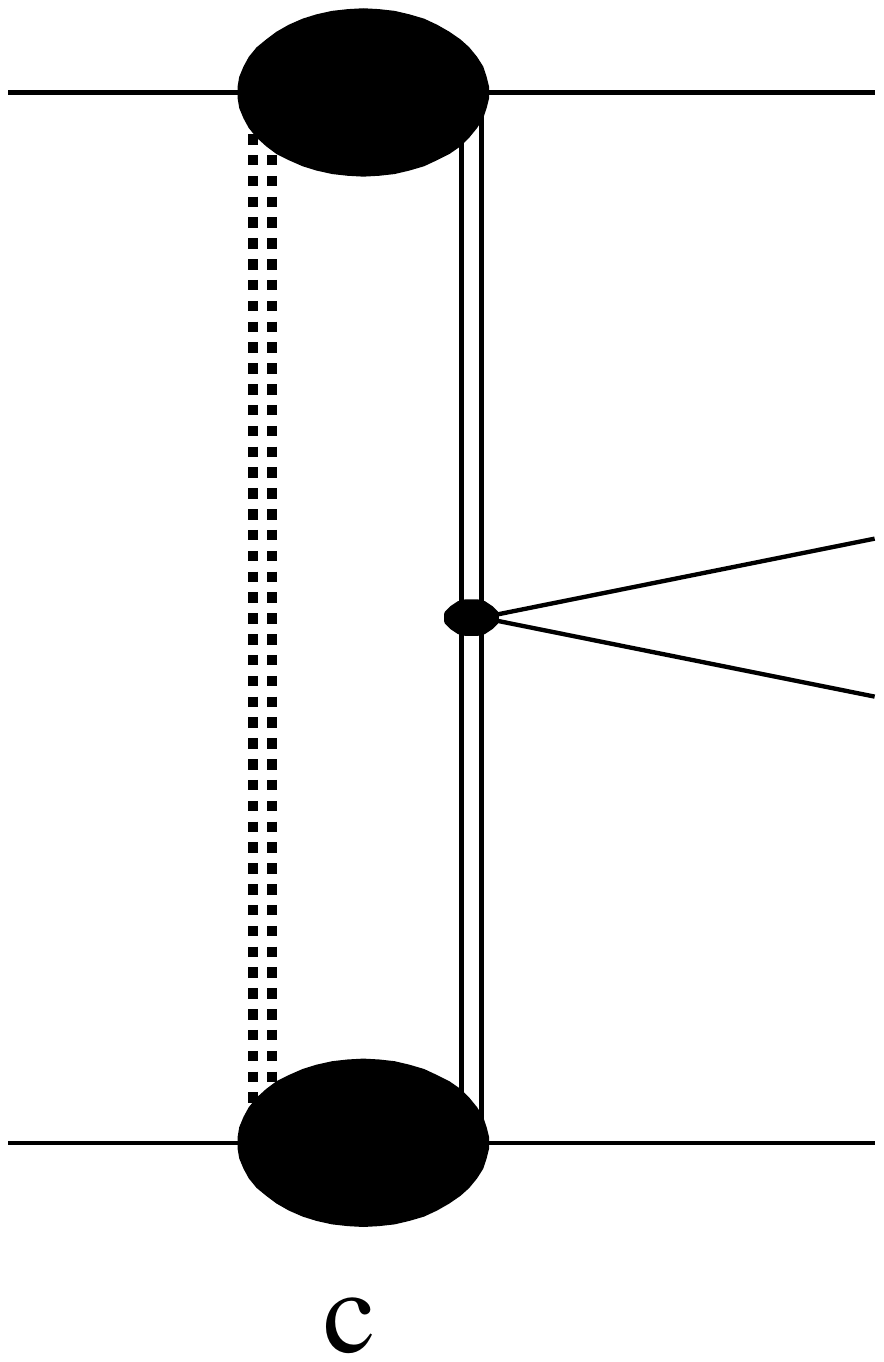}
\vspace{-1.8cm}
\caption{\sf (a) Central Exclusive Production, (b) the diagram relevant when we discuss the survival probability of the LRG in Section \ref{sec:3}.}
\label{fig:2}
\end{figure}
In case of the maximal Odderon the real part of amplitude $A(Y)$ grows as
Re$A=c\ln^2 s=cY^2$. On the other hand, the $t$-slope $B\propto R^2$, with the 
interaction radius limited by the Froissart~\cite{Fr} condition $R\leq {\rm const}\cdot Y$. 
That is 
the integral
\begin{equation}
I=\int dt |A(Y,t)|^2\sim Y^2 
\end{equation}
leading to
\begin{equation}
\sigma^{\rm CEP}~=~N\int_0^Y dy |I(y)\cdot V^2\cdot I(Y-y)|~\propto ~Y^5\ .
\end{equation}
Thus in such a case, the CEP cross section would grow much faster than the total cross section 
$\sigma_{\rm tot}\sim \ln^2s~=~Y^2$.

The same result can be obtained in impact parameter, $b$, space. Now
\begin{equation}
\label{b1}
\sigma^{\rm CEP}=N'\int_0^Ydy\int d^2b_1 d^2b_2 |A(y_1,b_1)\cdot V\cdot 
A(Y-y,b_2-b_1)|^2~\propto ~Y^5\ .
\end{equation}
Recall that in $b$ space the amplitude is limited\footnote{This is illustrated in Fig.4 below in terms of the partial wave amplitude $a_l(s)$.  The plot shows $|a_l|\le 2$.} to $|A(Y,b)|\leq 2$ by the 
unitarity equation
\begin{equation}
\label{un1}
2\mbox{Im}A(Y,b)=|A(Y,b)|^2+G_{\rm inel}(Y,b)
\end{equation}
where $G_{\rm inel}$ denotes the total contribution of all the inelastic channels. 
On the other hand the area where the amplitude is large ($A\sim O(1)$), that 
is the value of $\int d^2 b\sim \pi R^2\propto Y^2$, increases as 
$R^2\sim Y^2$.

Summing up the analogous cross sections for processes with a larger number of LRGs 
(i.e. a larger number, $n$, of hadron groups $X_i$ in Fig.1) we obtain the cross section 
which increases faster than the power of $s$. Indeed, each additional gap brings a factor
$\ln s$ arising from the integral over the gap size (times the `elastic' cross 
section which in the Froissart limit increases as $\ln^2 s_{i,i+1}$). The sum of these 
$\ln s$ factors leads to the power behaviour.

Note that by working in $b$ space we have a stronger constraint since for each value of 
$b$, that is for each partial wave $l=b\sqrt s/2$ of the incoming
 proton pair, the `total' cross section, $\sigma(b)_{\rm tot}$ must be less than the 
 corresponding CEP contribution.  
 
Actually one will face this FK problem in any model where the elastic cross section does 
not decrease with energy. 

At first sight the simplest way to avoid the FK problem is to say that the production 
vertex ($V$ in Fig.1) vanishes, at least as $t_i\to 0$. However this cannot be 
true. Indeed, as far as we have a non-vanishing high-energy elastic proton-proton cross section, we can build the diagram on the right side of Fig.\ref{pp} from a lower part which is just elastic $pp$-scattering and an upper part which corresponds
to the proton-antiproton elastic interaction. Such a diagram is generated by the $t$-channel two-particle unitarity equation for the amplitude,
\begin{equation}
\mbox{disc}_t~{\bf A}_{12}~=~\sum_j{\bf A}_{1j}^*|j\rangle\langle j|{\bf A}_{j2}\;\; ,
\end{equation}
where in our case $|j\rangle$ is the $t$-channel $p\bar p$ state.
%\vspace{-2cm}
\begin{figure} %[h]
%\begin{center}
\vspace{-4cm}
\includegraphics[scale=0.5]{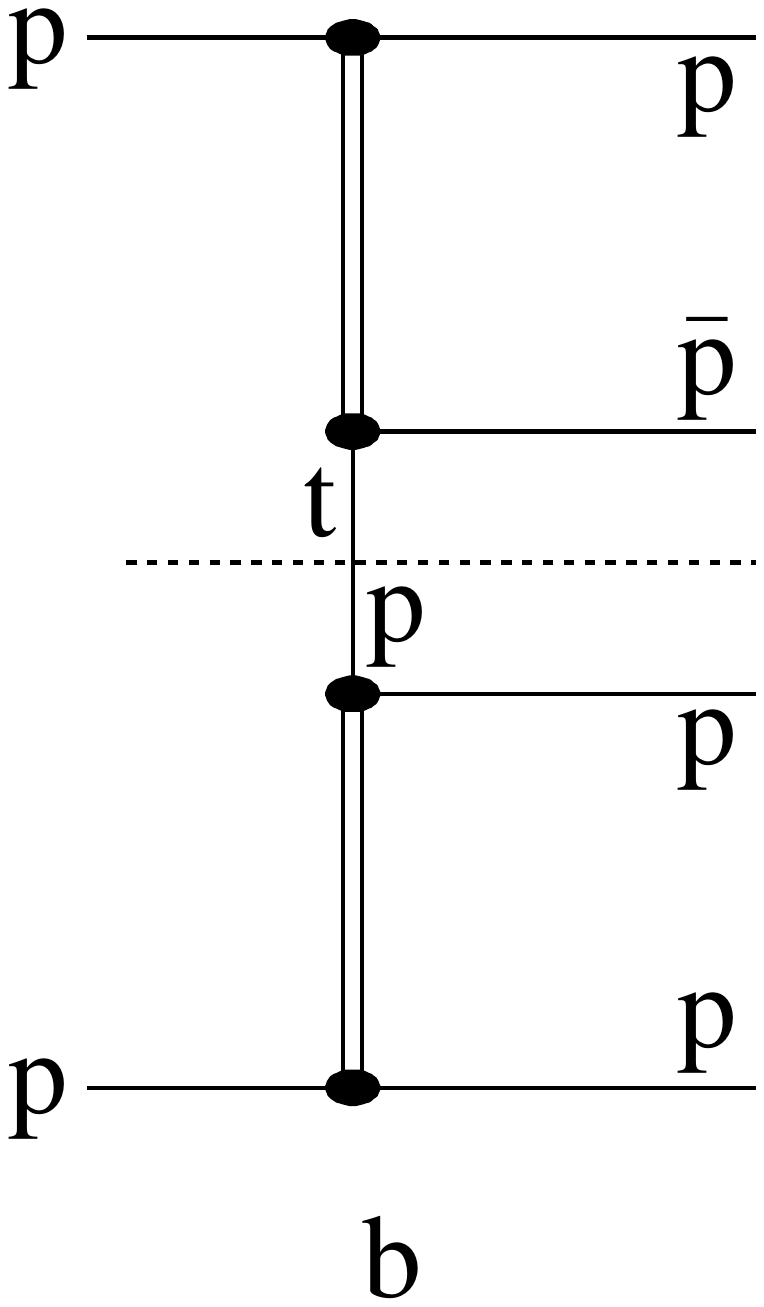}
\hspace{-2cm}
\includegraphics[scale=0.5]{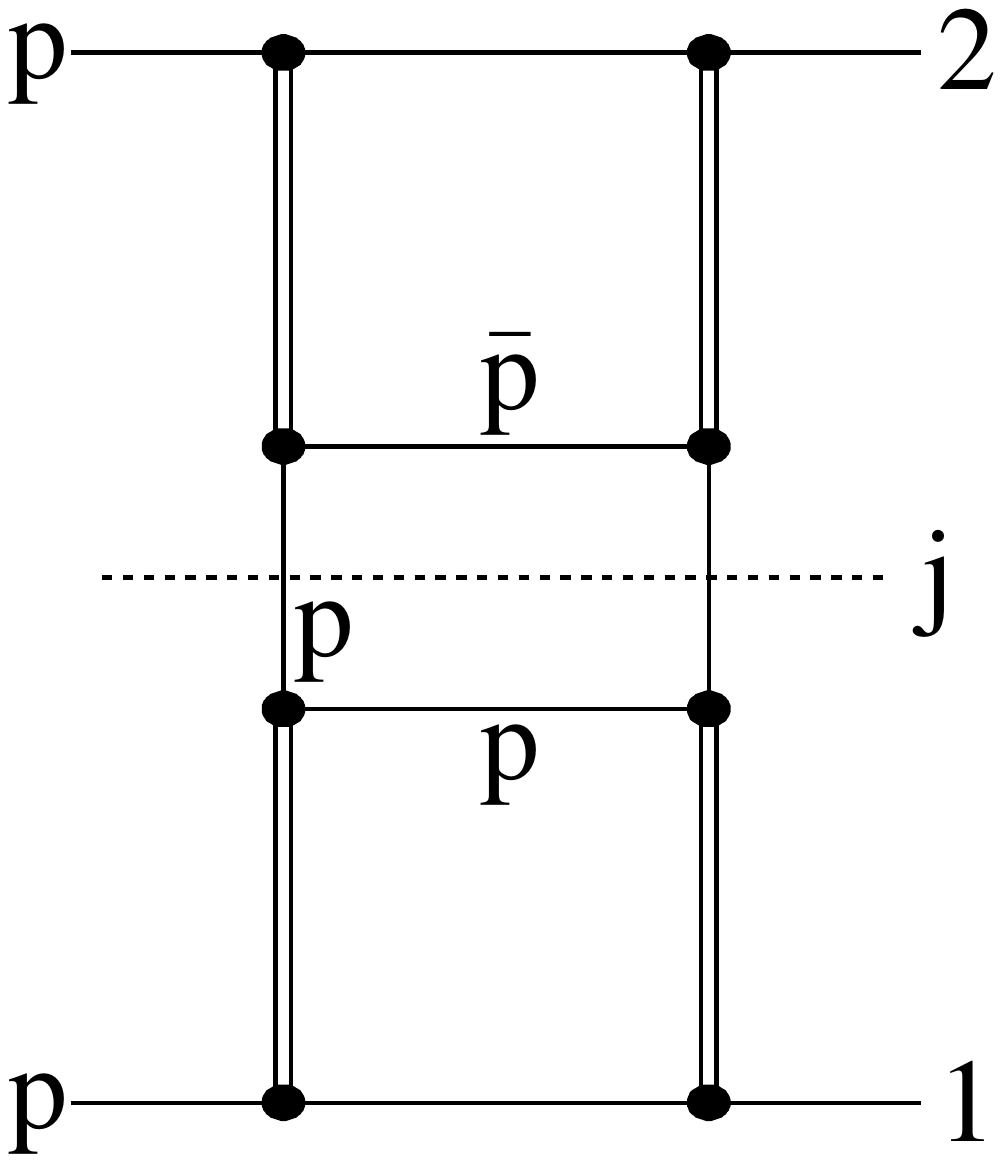}
\vspace{-2.5cm}
\caption{\sf Diagrams for the amplitude (left) and the cross section (right) of $p\bar p$ exclusive production generated by $t$-channel unitarity.}
\label{pp}
%\end{center}
\end{figure}
% In particular, $t$-channel unitarity via the cut, shown in Fig.2b by the 
%dashed line, generates a 
Note that the contribution of this diagram
 is singular at $t=m^2_p$ (where $m_p$ is the proton mass). There are no other
 similar terms corresponding to the central exclusive production of a $p\bar p$ pair with the same pole singularity. That is, in the vertex $V$ of Fig.2a, there exists at least one subprocess ($p\bar p$ CEP), which  
cannot be  cancelled identically.

It is useful to clarify the above argument, since it is a little subtle.
We have to distinguish between the momenta transfer squared, 
$t_1$ and $t_2$, incoming to the vertex $V$ in Fig.2a, and the momentum transferred squared inside 
the vertex $V$, denoted by $t$ in Fig.3a.
The value of the $t$ is driven by the transverse momentum, $p_t$, of the
antiproton.
Even if, due to a subtraction in dispersion relation in $t$ that reconstructs the amplitude, we find at some $p_t$
point that $V=0$, this will not insure that the total vertex contribution vanishes. We will have
$V\neq 0$ at other $p_t$ values.
Now, to calculate the total CEP cross section, we have to integrate
over all available $p_t$, so finally we obtain a non-zero contribution of
this particular $p\bar p$ subprocess.

\section{The solution of the FK problem  \label{sec:3}}
The only known solution of this multi-Reggeon problem comes from `black disk' asymptotics 
of the high energy cross sections. In such a case the (gap) survival probability, $S^2$, 
of the events with a LRG, tends to zero at $s\to\infty$, and the
value of $\sigma^{\rm CEP}$ does not exceed $\sigma_{\rm tot}$ (for a review of diffractive 
processes at the LHC see e.g.~\cite{KMRS}) .    

In other words, besides the contribution of Fig.2a, we have to consider the diagram of 
Fig.2b, where the double-dotted line denotes an additional proton-proton (incoming hadron) 
interaction. This diagram describes the absorptive correction to the original CEP process, 
and has a negative sign with respect to the 
amplitude $A^a$ of Fig.2a. Therefore to calculate the CEP cross section we have 
to square the full amplitude
\begin{equation}
\label{ful}
|A_{\rm full}(b)|^2~=~|A^a(b)-A^b(b)|^2~=~S^2(b)\cdot |A^a(b)|^2\ ,
\end{equation}
 where
 \be
 S^2(b)=|e^{-\Omega(b)}|\ , ~~~~~{\rm with} ~~~~~{\rm Re}\Omega \ge 0 \ .
 \ee

 Indeed, in terms of {\bf S}-matrix, the elastic component $S_l=1+iA(b)$, and the unitarity equation (\ref{un1})
reflects the probability conservation condition 
\be
\sum_n{\bf S}^*_l|n\rangle\langle n|{\bf S}_l~=~1
\ee
for the partial wave $l=b\sqrt s/2$. The 
solution of unitarity equation (\ref{un1}) reads
\begin{equation}
\label{el}
A(b)=i(1-e^{-\Omega(b)/2})\ ,
\end{equation}
or in terms of the partial wave amplitude with orbital moment $l=b\sqrt s/2$
\be
a_l~=~i(1-e^{2i\delta_l})~=~i(1-\eta_le^{2i{\rm Re}\delta_l})
\ee
where 
\be
\eta_l=e^{-2{\rm Im}\delta_l}~~{\rm with}~~~ 0\le \eta_l \le 1.
\ee
The unitarity circle bounding the partial wave amplitude is shown in Fig.~\ref{fig:f2}.
\begin{figure} [h]
\begin{center}
\includegraphics[clip=true,trim=0.0cm 0.0cm 0.0cm 0.0cm,width=12.0cm]{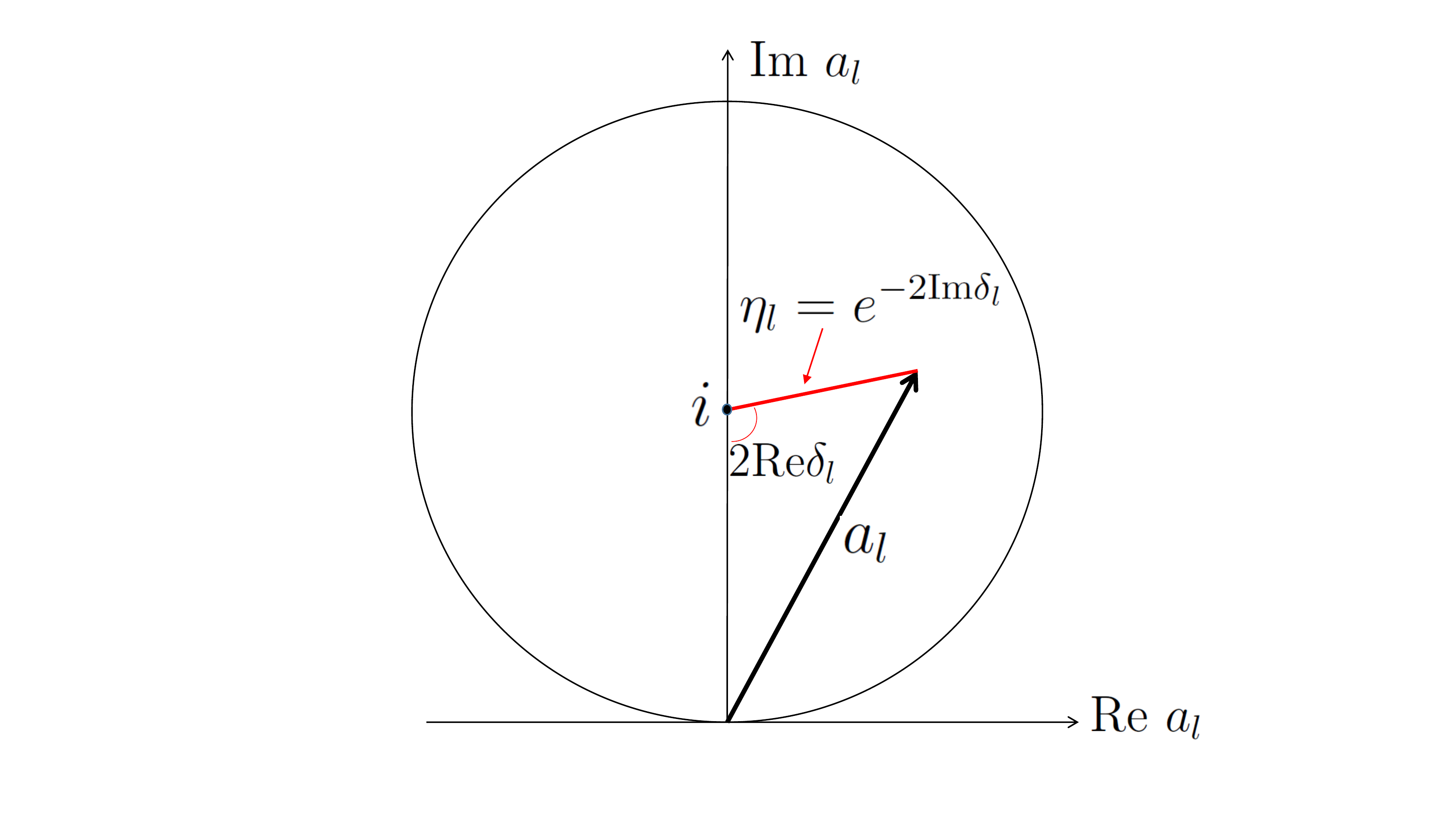}
%\vspace{-7.5cm}
\caption{\sf The partial wave amplitude, $a_l~=~i(1-e^{2i\delta_l})$, is constrained to lie in the `unitarity circle' centred on $(0,i)$. At low energy, for a resonance of orbital momentum $l$ appearing only in the elastic channel, the energy dependence of the amplitude $a_l$ follows a counter-clockwise circle of radius $\eta_l=1$ with maximum amplitude at $\delta_l=\pi/2$. However, for high energy scattering, black disk asymptotics requires Re$\Omega(b)\to \infty$ (see (\ref{eq18})), so 2Im$\delta_l\to \infty$ and $\eta_l\to 0$, and hence
 the amplitude becomes pure imaginary.}
\label{fig:f2}
\end{center}
\end{figure}

% where $-\Omega/2=2i\hat\delta_l=-\eta+2i\delta_l$ (where $\delta_l=\mbox{Re}\hat\delta_l$ and $\eta=\mbox{Im}2\hat\delta_l$) plays the role of scattering phase $\delta_l$ in the partial
 %wave analysis for the component with the orbital moment $l=b\sqrt s/2$; i.e. the elastic 
 The above discussion shows that $-\Omega(b)/2$ plays the role of $2i\delta_l$. The elastic component of {\bf S} matrix  $S_l=\exp(2i\delta_l)$.
 % with $\eta_l=1$. 
 Correspondingly, the probability of inelastic interaction (with all the intermediate states
 $n'$ except of the incoming, elastic, state) 
 \be
 G_{\rm inel}~=~\sum_{n'}{\bf S}^*_l|n'\rangle\langle n'|{\bf S}_l
 \ee
 takes the form 
 \begin{equation}
 \label{inel}
 G_{\rm inel}(b)~=~1-S^*_lS_l~=~1-e^{-{{\rm Re}\Omega(b)}}\ .
 \end{equation}
 Within the eikonal model $\Omega(b)$ is described by the sum of single Reggeon exchanges 
 while the decomposition of the exponent generates the multi-Reggeon diagrams.
 
 The gap survival factor, $S^2$, is the probability to observe a pure CEP event, where the 
 LRG is not populated by secondaries 
 produced in an additional inelastic interaction shown by the 
 dotted lines in Fig.2b. That is according to (\ref{inel})
 \begin{equation}
 \label{gap1}
 S^2(b)~=~1-G_{\rm inel}(b)~=~e^{-{\rm Re}\Omega(b)}\ .
 \end{equation}
 Equation (\ref{gap1}) can be rewritten as (see (\ref{el},\ref{inel}))
 \begin{equation}
 \label{gap2}
 S^2(b)=|1+iA(b)|^2=|S_l|^2\ .
 \end{equation}
 
 In the case of {\it black disk} asymptotics\footnote{Recall that the word `black' means the {\em complete} absorption of the incoming state (up to power of $s$ suppressed corrections). That is, Re$\Omega(s,b)\to \infty$. `Black disk' means  
 that in some region of impact parameter space, $b < R$, the whole initial 
wave function is absorbed.
  That is, the value of $S(b) = 1 + iA(b) = S_ l \to 0$, i.e. $A(b) \to i$.}
 \be 
 {\rm Re}\Omega(b)\to\infty ~~{\rm and}~~ A(b)\to i,
 \label{eq18}
 \ee
 for $b<R$.   That is, we get $S^2(b)\to 0$. The decrease of the gap survival probability 
 $S^2$ overcompensates the growth of the original CEP cross section (Fig.2a), so that 
 finally we have no problem with unitarity.
 
 Recall that this solution of the FK problem was actually realized by Cardy 
 in~\cite{Cardy}, where the reggeon diagrams (generated by Pomerons with intercept 
 $\alpha_P(0)>1$) were considered by assuming analyticity in the number of Pomerons;
  and also by Marchesini and Rabinovici in~\cite{MR}, where diffractive production 
  processes were discussed for the case of $\alpha_P(0)>1$.

  Note that at the moment we deal with a one-channel eikonal.   
  In other words, in Fig.2 and in the unitarity equation (\ref{un1}), we only account for the pure 
  elastic intermediate states 
  (that is the proton, for the case of $pp$ collisions). In general, there may be 
  $p\to N^*$ excitations shown by the black blobs in Fig.2b.
  The possibility of such  excitations can be included by the  
  Good-Walker~\cite{GW} formalism in terms of G-W eigenstates, $|\phi_i\rangle$,  which 
  diagonalize   the high energy scattering process; that is 
  $\langle\phi_k|A|\phi_i\rangle=A_k\delta_{ki}$. 
  In this case we encounter the FK problem for each state $|\phi_i\rangle $, and we 
  then solve 
  it for the individual eigenstates.

  \subsection{Edge of the disk}
  This subsection is not crucial for our final result, but it should be mentioned in order 
  to demonstrate the self-consistency of the whole picture.
  
 While the survival factor $S^2$ solves the FK problem for the central part of the black 
 disk, we still have to address the question of what happens at the edge of the disk, where 
 the optical density is not large? 
 That is, when $\mbox{Re}\Omega(b)\sim O(1)$. For large partial waves, 
 which occur in this domain, we still 
 may have CEP (and other diffractive LRG) cross sections larger than the total cross 
 section corresponding to such $l$-waves.
 
 The solution is provided by the fact that actually the constraint on the interaction 
 radius $R$ is a bit stronger than just $R\leq c \ln s$.
 %\lapproxeq c \ln s$.
 It was shown in~\cite{LR,Az} that we have to account for the `$\ln\ln s$' correction 
 \begin{equation}
 \label{LL}
 R~=~c\ln s ~-~\beta\ln\ln s=cY-\beta\ln Y\ .
 \end{equation}
 In this case, the radius of the CEP interaction shown in Fig.2a, (that is, before we 
 account for the screening effects of Fig.2b),
 \begin{equation}
 R_{\rm CEP}=R(y)+R(Y-y)=cY-\beta\left[\ln y +\ln (Y-y)\right]<R(Y)=c Y-\beta\ln Y,
 \end{equation}
 turns out to be smaller than that corresponding to elastic scattering. That is
 the multi-Reggeon amplitude is placed {\em inside} the black disk, and its contribution is strongly suppressed by the $S^2(b)$ factor.
 
 It was shown in~\cite{LR,DKT} that the same condition (\ref{LL}) provides the 
 possibility to satisfy the $t$-channel unitarity.

 \section{Maximal Odderon}
 Now let us consider the situation with the maximal Odderon, where at very high energies 
 the real part of elastic amplitude $A$ is comparable with its imaginary part, in the 
 sense that the ratio\footnote{Recall that the dotted 
 lines in Fig.2b denote just the elastic amplitude of proton-proton, or G-W 
 eigenstate, scattering.}
 \be
 {\rm Re}A/{\rm Im}A ~~\to ~~{\rm constant}\neq 0.
 \ee
  In such a case the elastic amplitude $A(b)$  has a non-zero real part, which violates the condition  $A\to i$ at $s\to\infty$. Now
the survival factor (\ref{gap2})
\be
S^2~=~|1+iA|^2~\geq ~|{\rm Re}A|^2~\neq ~ 0  
\ee
 tends to some non-zero constant.
 That is we loose the possibility to compensate the growth of the multi-Reggeon (CEP) 
 cross sections by the $S^2$ factor. Thus these cross sections, which increase faster 
 than the total cross section, will violate unitarity.
 
 Let us consider a dynamical model.
 Note that the expression for the elastic amplitude (\ref{el}) is an {\em exact} solution of the  $s$-channel two-particle unitarity equation (\ref{un1}), where 
 $\Omega$ is the two-particle-irreducible amplitude which includes all possible inelastic interactions.  That is,
  in terms of Regge theory, the Odderon contribution must be included into the opacity, and the opacity $\Omega(b)$ (i.e. the `phase' $\delta_l$) should be written at high energies as the sum of the even-signature (Pomeron) and the odd-signature (Odderon) terms
 \begin{equation}
 \Omega(b)~=~-i\left[{\rm Pomeron}(b) + {\rm Odderon}(b)\right]\ ,
 \end{equation}
 where the Pomeron term is mainly imaginary while the  Odderon contribution is mainly real.
 If $\alpha_P(0)=1+\Delta>1$, then the Pomeron term increases as the $s^\Delta$. 
 That is the exponent $\exp(-\Omega/2)\to 0$ and the second term in elastic 
 amplitude (\ref{el}) vanishes {\em together with the Odderon}  contribution.
  In other words, in the black disk limit when the value of $\mbox{Re}\Omega$ increases and
  $\exp(-\Omega/2)\to 0$ the Odderon contribution dies out.  
 The only chance to have a sizeable Odderon as $s\to\infty$ is to collect the 
 contribution from the edge of black disk where  the opacity 
 $\mbox{Re}\Omega(b)\sim O(1)$ is not large. From this region one may get an Odderon 
 contribution to the forward elastic amplitude
  $A_{\rm Odd}(t=0)\propto \ln s$; that is Re$A_{\rm Odd}$ could grow as the area of the ring 
  around the black disk, but certainly it cannot increase as $\ln^2 s$.
  
   However even the $\ln s$ asymptotic behaviour is questionable. The point is that the 
   radius of the Odderon induced interaction is most probably smaller than the radius 
   of black disk, generated by the even-signature bare Pomeron. Indeed, the nearest 
   $t$-channel singularity of the even-signature amplitude is $t=4m^2_\pi$, while 
   for the Odderon the nearest singularity is at $t=9m^2_\pi$. We cannot 
   build the Odderon state from two pions. The most reasonable appropriate hadron 
   state in the Odderon channel is the $\omega$ meson.    Therefore the growth of the Odderon radius with 
   energy is expected to be less than the growth of the black disk radius driven 
   by the even-signature amplitude, and the whole Odderon contribution will 
   be `absorbed' (i.e. power of $s$ suppressed) by the black disk.
 
 Thus in this section we have demonstrated that
 
 a) the maximal Odderon violates multiparticle $s$-channel unitarity
 
 b) the Odderon contribution disappears in the black disk limit when
  $\mbox{Re}\Omega\to \infty$.~\footnote{A similar conclusion was reached  in~\cite{FTan} based on the eikonal model which includes both the bare Pomeron and the bare Odderon poles. It was shown that, even starting with the Odderon with a larger than one intercept  (i.e. $\Delta_{\rm Odd}=\alpha(0)_{\rm Odd}-1>0$), after  eikonalization we get an Odderon contribution ($\sim s^{-\delta}$) which decreases as a power of energy, except for the case when the bare Odderon  trajectory coincides with the bare Pomeron trajectory.}

 \section{Reflective scattering}
 The same argument can be used to reject the so-called `reflective scattering' 
 asymptotics proposed in~\cite{TT1}. 
 % (see also~\cite{TT2,Drem,anis}). 
  Indeed, in 
 this regime it is assumed that the high energy interaction becomes pure elastic and the amplitude
 \be
 A(b<R)\to 2i~~~{\rm as} ~~~ s\to\infty,
 \ee  
(with our normalization %given in 
 fixed by eq.(\ref{un1})). This means that at very high energies we 
 will have an almost pure elastic interaction with $G_{\rm inel}\to 0$. In such a case 
 $S^2=1$ (see (\ref{gap2})).
 
On the other hand, $t$-channel unitarity generates the inelastic CEP  diagram Fig.2b with a cross section which increases faster than the elastic cross section. The contribution of such a diagram cannot be suppressed by  absorptive effects since now we have $S^2=1$.
  That is again we face the FK problem -- the cross section 
 of multi-Reggeon processes (in particular CEP) violates the unitarity 
 constraint.\\

We emphasize that black disk  absorption is the only cure of the FK disease. Thus any asymptotic behaviour of a high energy cross section, increasing with energy, which does not lead to complete absorption, is not consistent with multi-particle unitarity. 
In particular, the amplitudes  considered in ~\cite{TT2,Drem,anis}, should be abandoned, since they do not satisfy the black disk condition. 

 \section*{Acknowledgements}
 
VAK acknowledges  support from a Royal Society of Edinburgh  Auber award. MGR thanks the IPPP of Durham University for hospitality.

\end{document}